\documentclass{aa}
\pdfoutput=1
\usepackage{graphicx}
\usepackage{amsmath,amsfonts,amssymb}
\usepackage{color}
\usepackage[breaklinks,colorlinks,urlcolor=blue,citecolor=blue,linkcolor=magenta]{hyperref}
\usepackage{verbatim}
\usepackage{enumitem}
\usepackage{natbib}
\usepackage{aas_macros}
\bibpunct{(}{)}{;}{a}{}{,}

\usepackage{lineno}

\newcommand{\Msun}{M_\odot}
\newcommand{\td}{{\rm d}}
\newcommand{\vect}[1]{\boldsymbol{#1}}

\newcommand{\be}{\begin{equation}}
\newcommand{\ee}{\end{equation}}
\newcommand{\bea}{\begin{equation} \begin{aligned}}
\newcommand{\eea}{\end{aligned} \end{equation}}

\def\lsim{\mathrel{\raise.3ex\hbox{$<$\kern-.75em\lower1ex\hbox{$\sim$}}}}
\def\gsim{\mathrel{\raise.3ex\hbox{$>$\kern-.75em\lower1ex\hbox{$\sim$}}}}

\begin{document}

\title{Starlight from JWST: \\ Implications for star formation and dark matter models}
\author{
Juan Urrutia \inst{1,2}\fnmsep\thanks{juan.urrutia@kbfi.ee}
\and 
John Ellis \inst{3,4}\fnmsep
\and 
Malcolm Fairbairn \inst{3}\fnmsep
\and
Ville Vaskonen \inst{1,5,6}\fnmsep
}

\institute{
Keemilise ja Bioloogilise F\"u\"usika Instituut, R\"avala pst. 10, 10143 Tallinn, Estonia
\and
Department of Cybernetics, Tallinn University of Technology, Akadeemia tee 21, 12618 Tallinn, Estonia
\and
King’s College London, Strand, London, WC2R 2LS, United Kingdom
\and
Theoretical Physics Department, CERN, Geneva, Switzerland
\and
Dipartimento di Fisica e Astronomia, Universit\`a degli Studi di Padova, Via Marzolo 8, 35131 Padova, Italy
\and
Istituto Nazionale di Fisica Nucleare, Sezione di Padova, Via Marzolo 8, 35131 Padova, Italy
}

\abstract{
We confront the star formation rate in different dark matter (DM) models with UV luminosity data from JWST up to $z\simeq25$ and legacy data from HST. We find that a transition from a Salpeter population to top-heavy Pop-III stars is likely at $z\simeq10$ and that beyond $z=10-15$ the feedback from supernovae and active galactic nuclei is progressively reduced, so that at $z\simeq25$ the production of stars is almost free from any feedback. We compare fuzzy and warm DM models that suppress small-scale structures with the CDM paradigm, finding that the fuzzy DM mass $> 5.6 \times 10^{-22}{\rm eV}$ and the warm DM mass $> 1.5\, {\rm keV}$ at the 95\% CL. The fits of the star formation rate parametrisation do not depend strongly on the DM properties within the allowed range. We find no preference over CDM for enhanced matter perturbations associated with axion miniclusters or primordial black holes. The scale of the enhancement of the power spectrum should be $> 25\,{\rm Mpc}^{-1}$ at the 95\% CL, excluding axion miniclusters produced for $m_a < 6.6 \times 10^{-17}\,{\rm eV}$ or heavy primordial black holes that constitute a fraction $f_{\rm PBH} > \max[105 \Msun/m_{\rm PBH}, 10^{-4} (m_{\rm PBH}/10^4 \Msun)^{-0.09}]$ of DM. \\
~~\\
KCL-PH-TH/2025-12, CERN-TH-2025-085, AION-REPORT/2025-03}

\maketitle

\section{Introduction} 

An important target for James Webb Space Telescope (JWST) observations is to measure the UV luminosity function at high $z$~\citep{2022ApJ...940L..14N, Castellano:2022ikm, 2023MNRAS.518.6011D, 2023ApJ...946L..35M, 2023MNRAS.523.1036B, 2023ApJ...951L...1P, 2023ApJS..265....5H, 2024ApJ...960...56H, Perez-Gonzalez:2025bqr, Castellano:2025vkt}. These observations complement the more precise lower-$z$ observations from the Hubble Space Telescope (HST)~\citep{2016MNRAS.459.3812M, 2018ApJ...855..105O, 2021AJ....162...47B, 2022ApJ...928...52F, 2023MNRAS.524.5454L, 2024ApJ...961..209B}. The high-$z$ observations serve as a probe of galaxy dynamics and the star formation rate (SFR) at early epochs of the universe, constraining the abundance and masses of the conjectured first-generation Pop-III stars~\citep{Maiolino:2023wwm, 2023A&A...678A.173V, Ventura:2024mbp, Fujimoto:2025kbv}. The frontier of UV luminosity function measurements has been pushed back to unprecedented early times with recent measurements at $z\sim 25$, opening a window on the epoch before reionisation took place~\citep{Perez-Gonzalez:2025bqr, Castellano:2025vkt}. 

The number density of Pop-III stars depends on the abundance of early galaxies. This depends in turn on the nature of dark matter (DM), since galaxies populate DM halos. The UV luminosity function is, therefore, a valuable source of insights into DM models~\citep{Sabti:2021unj, Hutsi:2022fzw, Parashari:2023cui, Liu:2024edl, Winch:2024mrt, Sipple:2024svt}. As the redshift increases, the halo mass function (HMF) becomes more sensitive to the physical properties of DM, since fewer halos are built via hierarchical growth and the imprints from the physics before the CMB are more noticeable. 

Our understanding of structure formation and the star formation process in models of DM with suppressed small-scale structure formation at high $z \gtrsim 10$, such as Fuzzy Dark Matter (FDM) and Warm Dark Matter (WDM), or in models with an enhanced matter power spectrum, is very limited. Therefore, the very high redshift abundance and distribution of Pop-III stars, made of pristine gas from the Big Bang, are topics of hot debate. The roles of external radiation, internal fragmentation of the gas clouds and the dynamics that form the stars are highly uncertain, so the environments where these first stars were born and their mass function are as yet unknown; see~\citet{Klessen:2023qmc} for a review. The UV luminosity function measured with JWST has proven to be higher than anticipated, even when considering only the galaxies which are spectroscopically confirmed~\citep{2024ApJ...960...56H}. This suggests that either the feedback mechanisms preventing star formation were less restrictive~\citep{2004ApJ...600....1S, 2023MNRAS.523.3201D, 2021MNRAS.506.5512F}, or that the Pop-III stars were very top-heavy and therefore more luminous~\citep{2015MNRAS.448..568H, 2023ApJ...951L..40S, 2014ApJ...781...60H}. This tension has been exacerbated by the measurements of starlight at $z\sim25$, which are higher than the predictions from many semi-analytical models in the literature~\citep{Perez-Gonzalez:2025bqr}. 

The objectives of this work are to parameterise the essential physics of high-$z$ star formation and understand the features necessary to reproduce the UV luminosity function back to $z\sim25$ in different DM models. We take into account possible degeneracies and features of modified DM models, feedback and changes in the stellar population. To this end, we connect the HMF obtained using the excursion set formalism~\citep{Bond:1990iw}, with the early Universe microphysics of DM, and combine this with a phenomenological parameterisation of star formation. We also discuss the effects of gravitational lensing and dust attenuation, neither of which seems to be very significant for the interpretation of the $z>15$ UV data.

To extrapolate the results of DM simulations from low- to high-$z$ and get reliable bounds, we improve on existing techniques in the literature in two key aspects: 1) We derive a new estimate of the DM halo growth rate that accounts for the ellipsoidal collapse correction, and use it consistently for different DM models. To this end, we recompute the first-crossing probability distributions of the paths with a moving barrier and find a better fit than that originally proposed by~\citet{Sheth:1999su}. The conditional mass function in the Sheth-Tormen approach is divergent for very small steps, as has been pointed out by~\citet{Jiang:2013kza} and \citet{Zhang:2008dx}. Our fit is divergence-free and reproduces analytical estimations of~\citet{Zhang:2008dx}. 2) We provide new values for the parameters the smooth-$k$ window function that, unlike those proposed in~\citet{Leo:2018odn}, match consistently the CDM prediction.

We find that, under the assumption that the star formation rate (SFR) is proportional to the halo growth rate, the observations of the UV luminosity function up to high $z$ indicate two interesting effects that change the luminosity function. First, at $z > 10$, either the stars need to be more luminous and/or the SFR needs to be higher. This transition is relatively sharp and might be caused by a top-heavy Pop~III-dominated star population. We find that the evidence for this effect above the $95\%$ CL. In addition, the SFR needs to be further enhanced at $z>15$. This can be achieved by reducing the feedback from supernovae (SNe) and active galactic nuclei (AGNs) on the SFR. We find that to explain the $z>15$ observations, these feedbacks need to be significantly reduced to match the UV luminosity function, and at $z \simeq 25$ the production of stars should be almost free from feedback effects. {In agreement with~\citet{Yung:2025ttv}, we find that, with these changes in the stellar population and feedback effects, the JWST data can be explained within the CDM model.}

An important focus of this work is whether different DM scenarios require different star formation physics to fit high-redshift data.  Interestingly, we find very little sensitivity to DM properties that differ from CDM, with only a small change in star formation rates favoured for WDM, and no change for FDM.

When marginalizing over the SFR parametrization, we find a lower bound of $m_{\rm FDM} > 5.6 \times 10^{-22}{\rm eV}$ for the case of FDM and $m_{\rm WDM} > 1.5\, {\rm keV}$ for the case of WDM. The new JWST data, although pushing the observations to much earlier galaxies, are still not precise enough to put much stronger bounds than the HST data, as also found in~\citet{Winch:2024mrt}. We also find that enhanced small-scale matter perturbations are not preferred over changes in the star formation due to the presence of Pop-III stars and changes in the feedback in the pre-reionisation era. We are able to put bounds on the scale of white-noise enhanced perturbations at $k_c > 25\,{\rm Mpc}^{-1}$, excluding, for example, the possibility of axion miniclusters produced for $m_a < 6.6 \times 10^{-17}\,{\rm eV}$ or heavy primordial black holes (PBHs) that constitute the fraction $f_{\rm PBH} > \max[105 \Msun/m_{\rm PBH}, 10^{-4} (m_{\rm PBH}/10^4 \Msun)^{-0.09}]$ of DM. In a forthcoming paper, we will leverage these findings to compute the growth of SMBHs in non-CDM scenarios and obtain, for the first time, even more competitive bounds on deviations from CDM using the measurements of high-$z$ SMBHs by JWST. 

Throughout this work, we use the Planck 2018 CMB best-fit values for the cosmological parameters~\citep{Planck:2018vyg}: $\Omega_{\rm M} = 0.315$, $\Omega_{\rm B} = 0.0493$, $z_{\rm eq} = 3402.0$, $\sigma_8 = 0.811$, $h = 0.674$, $T_0 = 2.7255\,{\rm K}$ and $n_s = 0.965$.

\section{Halo mass function and growth rate}

\begin{figure*}
    \centering
    \includegraphics[width=\textwidth]{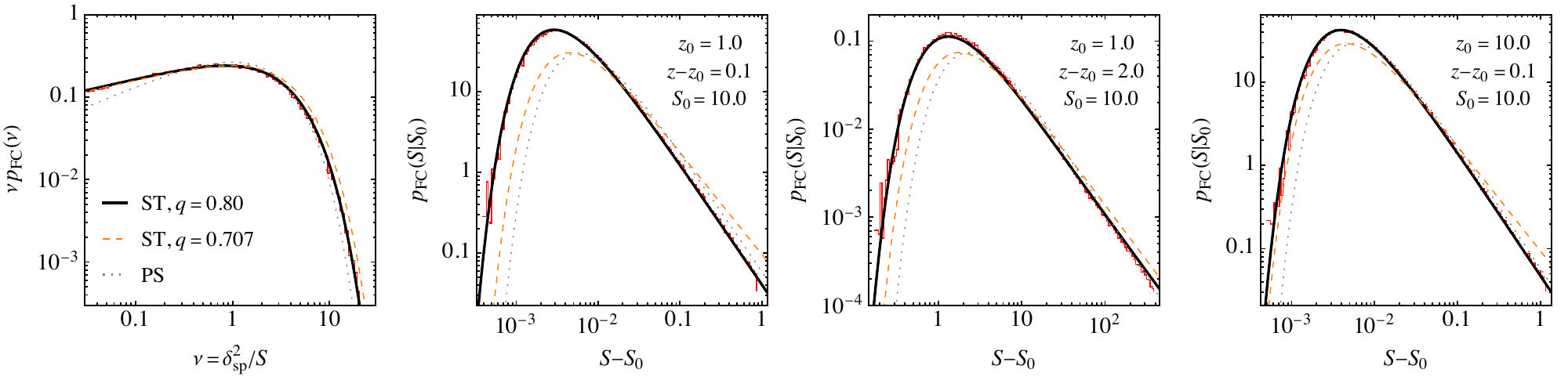}
    \caption{\textit{Left panel:} The unconditional first-crossing distribution for walks starting from $S = 0$, $\delta = 0$. The histogram shows the results of random walks, the solid black and dashed orange curves show the ansatz~\eqref{eq:ST} with $q=0.80$ and $q=0.707$, respectively, and the dotted grey curve shows the spherical collapse result. \textit{Other panels:} The conditional first-crossing distribution at $z>z_0$ for walks starting from $S_0>0$ and $\delta_{\rm ell}(z_0,S_0) > 0$ as indicated in the plot. The histogram again shows the results of random walks, the solid black curves show the ansatz~\eqref{eq:ST2}, the dashed orange curves show the ansatz~\eqref{eq:ST} with $\delta_{\rm sp}(z) \to \delta_{\rm sp}(z)-\delta_{\rm sp}(z_0)$ and $S \to S-S_0$, and the dotted gray curve shows the spherical collapse result.}
    \label{fig:walks}
\end{figure*} 

In the excursion-set formalism~\citep{Bond:1990iw}, the conditional and unconditional HMFs are obtained from the first-threshold-crossing distributions of uncorrelated random walks in the $(S,\delta)$ plane, where $S = \sigma(M)^2$ denotes the variance of matter perturbations at mass scale $M$ and $\delta$ is the matter density contrast around a given spatial point. The unconditional mass function is obtained from the first-crossing distribution $p_{\rm FC}(S,z)$ of walks starting from $(0,0)$:
\be
    \frac{\td n(z)}{\td M} = \frac{\rho}{M} \left|\frac{\td S}{\td M} \right|  p_{\rm FC}(S,z) \,.
\ee
For spherical collapse in a universe with CDM, the density contrast threshold is $\delta_{\rm sp}(z) = 3(3 \pi/2)^{2/3}/(5D_{\rm g}(z))$~\citep{Dodelson:2003ft}, where the CDM linear growth function $D_{\rm g}(z)$ does not depend on $S$ and the first-crossing distribution $p_{\rm FC}(S,z)$ can be computed analytically. In the extension to include ellipsoidal collapse, the threshold becomes $S$-dependent: 
\be \label{eq:deltaell}
    \delta_{\rm ell}(S,z) = \sqrt{a} \,\delta_{\rm sp}(z) \left[ 1 + 0.485 \left(a \frac{\delta_{\rm sp}(z)^2}{S} \right)^{-0.615} \right]
\ee 
with $a = 0.707$~\citep{Sheth:1999su}, and $p_{\rm FC}(S,z)$ cannot be calculated analytically. The standard Sheth-Tormen ansatz for the first-crossing distribution with ellipsoidal collapse is~\citep{Sheth:1999mn} 
\be \label{eq:ST}
    p_{\rm FC}(S,z) = A \left[ 1 + \left(q \nu \right)^{-p}\right] \sqrt{\frac{q\nu}{2\pi}} \frac{e^{-q\nu/2}}{S}  , \,\, \nu = \frac{\delta_{\rm sp}(z)^2}{S} ,
\ee
where the normalization factor $A$ is given by $A = [1+2^{-p} \Gamma(1/2 - p)/\sqrt{\pi}]^{-1}$, $p=0.3$ and $q = a = 0.707$.\footnote{The spherical collapse model corresponds to $p = 0$, $q = 1$ and $A = 1/2$.} In the left panel of Fig.~\ref{fig:walks}, we show that random walks prefer a slightly higher value of $q = 0.80$. The histograms shown in Fig.~\ref{fig:walks} are obtained by generating $2\times 10^5$ random walks with step size $\Delta \nu/\nu = 10^{-4}$. {The Seth-Tormen halo mass function has been found to match well the results of numerical simulations~(see, e.g.,~\citet{2003MNRAS.346..565R,Zheng:2023myp}).}

For the conditional mass function, we study random walks starting from $S_0 > 0$ and $\delta_0 = \delta_{\rm ell}(S_0,z_0)>0$. The simplest ansatz for the first-crossing distribution $p_{\rm FC}(S,z|S_0,z_0)$, as considered, e.g., by~\citet{Sheth:2001dp}, is obtained by the shift $\delta_{\rm sp}(z) \to \delta_{\rm sp}(z)-\delta_{\rm sp}(z_0)$ and $S \to S-S_0$ in~\eqref{eq:ST}. As seen in Fig.~\ref{fig:walks}, comparing the orange dashed curves with the histograms, we find that this ansatz does not reproduce well the results of the random walks. A better estimate, shown by the solid black curves in Fig.~\ref{fig:walks}, is given by 
\bea \label{eq:ST2}
    &p_{\rm FC}(S,z|S_0,z_0) = A' \left[ 1 + \left(a \nu \right)^{-p}\right] \sqrt{\frac{\nu'}{2\pi}} \frac{e^{-\nu'/2} }{S-S_0}  \,, \\ 
    &\nu = \frac{\delta_{\rm sp}(z)^2}{S} \,, \quad \nu' = \frac{(\delta_{\rm ell}(z,S_0)-\delta_{\rm ell}(z_0,S_0))^2}{S-S_0} \,,
\eea
where, as in Eq.~\eqref{eq:deltaell}, $a=0.707$ and the normalization factor $A'$ in the limit $z\to z_0$ is
\be
    A' = \left[1+\left(a \frac{\delta_{\rm sp}(z_0)^2}{S_0}\right)^{-p}  \right]^{-1} \,.
\ee
In the limit $z\to z_0$, our formula~\eqref{eq:ST2} agrees well with the analytical estimate derived by~\citet{Zhang:2008dx}. However, our expression also matches the numerical results for large redshift differences $z-z_0\gg 0$.~\footnote{We note that the ellipsoidal collapse thresholds at different redshifts can intersect. This occurs, however, only at very small halo masses, $M<10^4 \Msun$, which are not relevant for our study.}

From the first-crossing probabilities, we can construct the probability density that a halo whose mass at $z'$ is $M'$ ends up being a part of a halo of mass $M>M'$ at $z<z'$:
\be
    \frac{\td \tilde P(M,z|M',z')}{\td M} = p_{\rm FC}(S',z'|S,z) \frac{p_{\rm FC}(S,z)}{p_{\rm FC}(S',z')} \left|\frac{\td S}{\td M} \right| \,.
\ee
This, in turn, directly gives the halo mass growth rate:
\bea \label{eq:Mdot}
    \frac{\td M}{\td z} &= \lim_{\Delta z \to 0} \int_M^{2M} \td M'\, \frac{M'-M}{\Delta z} \frac{\td P(M',z|M,z+\Delta z)}{\td M'} \\
    &= \sqrt{\frac{2}{\pi}} \, \frac{\td \delta_{\rm ell}(S(M),z)}{\td z} \left|\frac{\td S}{\td M} \right|^{-1}  \sqrt{S(M)-S(2M)} \,.
\eea
Unlike the estimate derived by~\citet{Correa:2014xma}, our estimate~\eqref{eq:Mdot} accounts for the correction from ellipsoidal collapse. Moreover, our estimate is based on the expected increase in the halo mass, while the estimate of~\citet{Correa:2014xma} is based on finding the expected mass of the main progenitor. However, we find that these estimates agree within a factor two across a broad range of halo masses and redshifts.

\section{Dark matter models} 

\subsection{Suppressed small-scale structures}

We consider two DM models that predict suppression of small-scale structures: 1) FDM where the DM is provided by coherent waves of an ultralight bosonic field with mass $m_{\rm FDM} = \mathcal{O}(10^{-20}\,{\rm eV})$ and 2) WDM where the DM is provided by semi-relativistic particles with masses $m_{\rm WDM}=\mathcal{O}({\rm keV})$. The changes in the matter power spectrum in these cases can be described by transfer functions $T_J(k)^2$, where $J$ labels the model. The resulting matter power spectrum is given by
\be
    \mathcal{P}_{\rm sup}(k) = T_J(k)^2 \ \mathcal{P}_{\rm CDM}(k) \,,
\ee
where $\mathcal{P}_{\rm CDM}(k)$ denotes the CDM matter power spectrum that we compute with the transfer function from~\citet{Eisenstein:1997ik}.

In the case of FDM, the speed of sound is modified by quantum pressure, so that $c_s\sim k/(4a^2m_{\rm FDM}^2)$. Consequently, the Jeans scale is non-zero (see e.g.~\citet{Marsh:2015xka}),
\be
    k_{\rm J} = \frac{66.5\,{\rm Mpc}^{-1}}{(1+z_{\rm eq})^{1/4}} \left[\frac{\Omega_{\rm FDM}\,h^2}{0.12}\right]^{\frac14}\left[\frac{m_{\rm FDM}}{10^{-22}\,{\rm eV}}\right]^{\frac12}\, ,
\ee
and perturbations at small scales $k \gtrsim k_{\rm J}$ are suppressed. This suppression can be estimated using an effective fluid approximation that is accurately matched to the solution of the Klein-Gordon equation~\citep{Hu:2000ke,Passaglia:2022bcr}. We use the fitting formula provided by~\citet{Passaglia:2022bcr}:
\be
    T_{\rm F} = \frac{\sin x^n}{x^n(1+B x^{6-n})} \,,\quad
    x = A \frac{k}{k_{\rm J}} \,,
\ee
where 
\be 
A = 2.22 m_{22}^{1/25+1/1000 \ln m_{22}} \; {\rm and} \; B = 0.16
m_{22}^{-1/20}
\ee
with $m_{22} \equiv m_{\rm FDM}/10^{-22} \,{\rm eV}$.
 
In the case of WDM, a thermal production mechanism is typically assumed, and the particles are lighter than typical CDM particles, remaining relativistic for a longer time. This implies that the free streaming length of WDM particles can be long, $\lambda_{\rm fs} \sim (m_{\rm WDM}/{\rm keV})^{-4/3} {\rm Mpc}$. In order to compute the resultant damping of the small-scale structures, it is necessary to track accurately the transition between relativistic and non-relativistic behaviour by solving the Boltzmann equations. This yields the fit~\citep{Bode:2000gq,Hansen:2001zv,Viel:2005qj}
\be
    T_{\rm W} = \left[1+(\alpha k)^{2\mu}\right]^{-5/\mu} \,,
\ee
with $\mu=1.12$ and
\be
    \alpha = 0.049\left[\frac{m_{\rm WDM}}{\rm keV}\right]^{-1.11} \left[\frac{\Omega_{\rm DM}}{0.25}\right]^{0.11} \left[\frac{h}{0.7}\right]^{1.22} h^{-1} {\rm Mpc}\,.
\ee

\subsection{Enhanced small-scale structures}

For the enhancement of small-scale structures, we also consider two cases: 1) an axion DM model that includes axion miniclusters and 2) a scenario where heavy primordial black holes (PBHs) constitute a fraction of the DM density. In both cases, we model the enhancement of the matter power spectrum as~\citep{Hutsi:2022fzw}
\be \label{eq:enh}
    \mathcal{P}_{\rm enh}(k) = \mathcal{P}_{\rm CDM}(k) + \mathcal{P}_{\rm CDM}(k_c) \,.
\ee
This parametrisation includes one parameter, the scale $k_c$ above which the power spectrum is dominated by a white noise contribution. We perform the analysis in terms of $k_c$, which is related to the parameters of the underlying DM models as discussed below.

In axion-like particle DM models, where the global $U(1)$ symmetry is broken after cosmic inflation, significantly small-scale density fluctuations arise due to the Kibble mechanism. These lead to the formation of axion miniclusters~\citep{Kolb:1993zz}. This scenario results in a white noise contribution to the matter power spectrum~\citep{Fairbairn:2017sil,Feix:2019lpo,Ellis:2022grh}. The resulting power spectrum can be approximated by Eq.~\eqref{eq:enh} with
\be \label{eq:kc_AMC}
    k_c \approx 3 h\, \mathrm{Mpc}^{-1} \left[\frac{m_a}{10^{-18}\,\mathrm{eV}}\right]^{0.6} \,.
\ee
where $m_a$ denotes the axion-like particle mass. The white noise contribution is cut at scales that re-entered the horizon when the axion began oscillating:
\be
    k_{\rm cut} \simeq 300\, \mathrm{Mpc}^{-1} \sqrt{\frac{m_a}{10^{-18}\,\rm{eV}}} \, .
\ee
However, in the relevant part of the parameter space, the cut is much beyond the enhancement scale, $k_c \ll k_{\rm cut}$. Notice also that the Jeans scale is larger than the cut-off scale, $k_J > k_{\rm cut}$.

In scenarios involving heavy PBHs there is a white noise contribution to the matter power spectrum~\citep{Inman:2019wvr,DeLuca:2020jug} with
\be
    k_c \approx 6 h \, \mathrm{Mpc}^{-1} \left[\frac{f_{\rm PBH} m_{\rm PBH}}{10^4 \Msun}\right]^{-0.4} \, .
\ee
where $f_{\rm PBH}$ is the DM fraction in PBHs, $m_{\rm PBH}$ is the PBH mass, and $\rho_{\rm DM}$ is the dark matter density. The cut-off is set by the average separation of PBHs~\citep{Hutsi:2022fzw}:
\bea
    k_{\rm cut}= 900 h \, \mathrm{Mpc}^{-1} \left[f_{\rm PBH} \frac{10^4 \Msun}{m_{\rm PBH}}\right]^{1/3} \, .
\eea
Below this scale, one expects roughly one PBH per corresponding comoving volume, and the seed effect~\citep{1983ApJ...268....1C,Carr:2018rid,Cappelluti:2021usg} becomes dominant. The condition $k_c < k_{\rm cut}$ is satisfied for $f_{\rm PBH} > 10^{-4} (m_{\rm PBH}/10^4 \Msun)^{-0.09}$.

\begin{figure}
    \centering
    \includegraphics[width=\columnwidth]{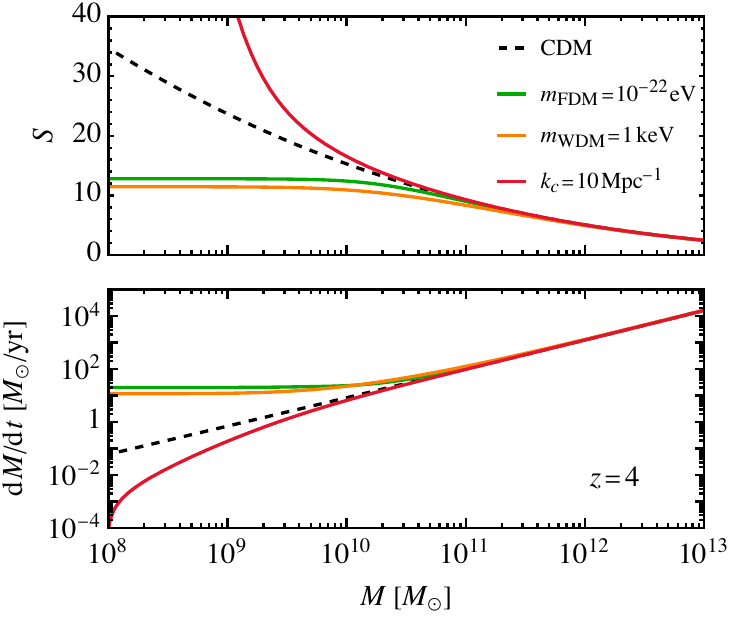}
    \caption{The variance $S=\sigma^2$ of matter perturbations (upper panel) and the halo growth rate (lower panel) in the models considered in this work: CDM (dashed black line), FDM (green line), WDM (orange line) and an enhanced matter power spectrum (red line).}
    \label{fig:dotM}
\end{figure}

\begin{figure*}
    \centering
    \includegraphics[width=0.9\textwidth]{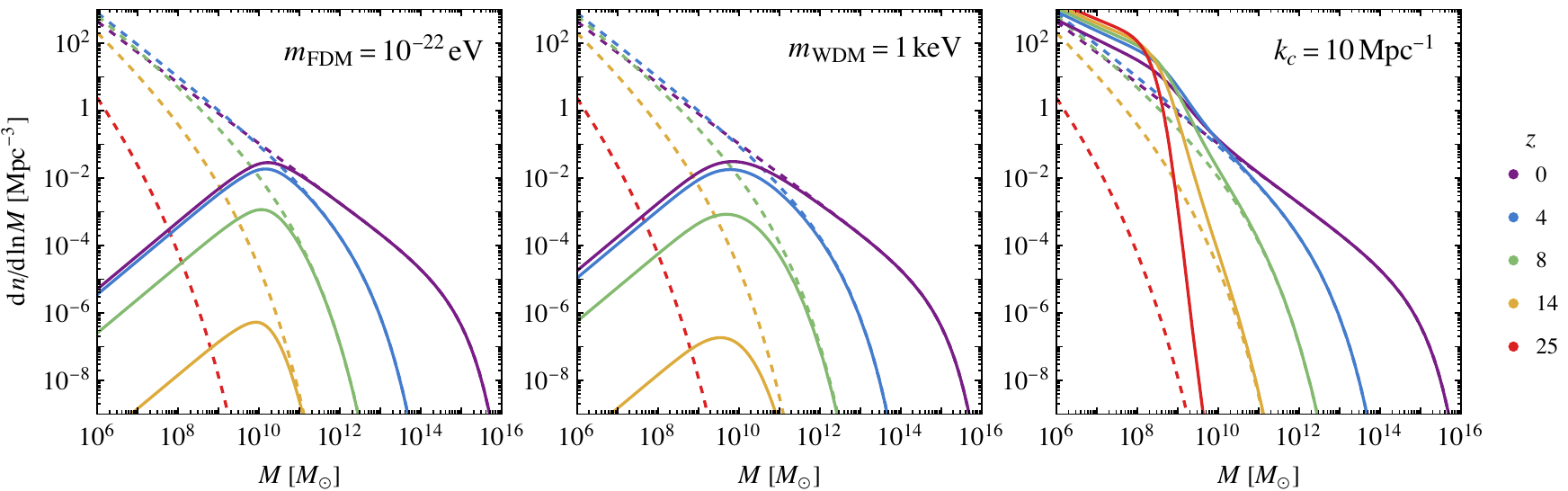}
    \caption{The HMF at various redshifts $z$ in the FDM (left) and WDM (middle) models, and in models where white noise dominates the matter perturbations at $k>k_c$ (right) . The dashed contours show for comparison the HMF in the CDM model.}
    \label{fig:HMF}
\end{figure*} 

\subsection{Window function}

In order to convert the matter power spectrum into the HMF, it is necessary to specify a window function to be used in conjunction with the excursion-set formalism. The window function is used to compute the variance of the matter perturbations, 
\be
    \sigma^2(R) = A\int\td k\, k^2 \mathcal{P}(k) \Tilde{W}^2(k|R) \,,
\ee
where the normalization factor $A$ is chosen to match the Planck measurement of $\sigma_8 = \sigma(R=8/h\,{\rm Mpc}) = 0.811$~\citep{Planck:2018vyg}. We use a window function of the form~\citep{Leo:2018odn,Verwohlt:2024efh}:
\be
    \tilde{W}(k|R) = \frac{1}{1+(c_1 k R)^{c_2}} \,,
\ee
where $R$ is related to the mass of the overdensity by $M = 4\pi \bar\rho R^3/3$. In the case of a suppressed matter power spectrum, the HMF scales as $\td n/\td \ln M \propto M^{c_2/6}$ and the halo growth rate as $\td M/\td z \propto M^{1-c_2/6}$ at scales much below the suppression scale. For all models, we use $c_2=6$ so that also for WDM and FDM models the growth rate increases with the halo mass, and fix $c_1 = 0.43$ so that the resulting variance in CDM matches well with that computed using the standard real space top hat window function, in agreement with simulation results. We find that this choice also reproduces well the simulation results in the FDM model~\citep{Schive:2015kza}, but it underestimates the halo mass where the HMF found in WDM simulations deviates from the CDM prediction~\citep{Lovell:2013ola}. On the other hand, the choice $c_1 = 1/3.3$ suggested by~\citet{Leo:2018odn} reproduces the low-mass part, but it does not match the CDM prediction at high masses. To reproduce the simulation results in WDM and to match CDM predictions at masses much above the suppression scale, we use $c_1 = 0.43$ in the window function and multiply $\alpha$ in the WDM transfer function by $3.3 c_1 \approx 1.4$.

We show in the upper panel of Fig.~\ref{fig:dotM} how the variance of the density perturbations in the models described above differs from that in the CDM model. For FDM and WDM, the variance asymptotes to a constant value at small masses. The break from the CDM-like behaviour at high masses is sharper for FDM than for WDM. This reflects the fact that the free streaming length does not make as abrupt a transition as the Jeans scale. For the case of enhanced perturbations, we see the opposite effect and the variance of the perturbation grows rapidly at small masses. In the lower panel of Fig.~\ref{fig:dotM}, we show the change in the DM halo growth rate for the different scenarios. In cases with suppressed small-scale structures, we find that the growth rate exceeds that of CDM, while the halos grow more slowly in the case with enhanced small-scale structures. 

The HMFs are shown in Fig.~\ref{fig:HMF}, where they are compared with the CDM predictions. For the case with suppressed small-scale structures, the HMF grows $\propto M$ at small masses. This power-law behaviour does not significantly affect the fits to the UV luminosity observations discussed below. For the enhanced perturbations, we can see how the enhancement gets weaker at low $z$ as the miniclusters merge with heavier halos and their original imprint is diluted.

\section{UV luminosity functions}

After discussing the DM structures, which provide the potential wells that trap baryons, we now turn to the source of the UV light, namely the stars that populate galaxies. The stellar mass in a galaxy increases not only via mergers, but mainly via the formation of new stars from cold gas. At high $z$, the star formation rate (SFR) can be probed using the UV luminosity function, which quantifies the number density of galaxies as a function of their UV luminosity. The UV luminosity of a galaxy is directly proportional to the SFR because the sources that dominate this emission are massive $\mathcal{O}(10\,M_{\odot})$ stars that are short-lived at cosmological scales, with $\tau \lesssim 10$\,Myr. \footnote{The timescale is further reduced if the Pop-III mass function is more top-heavy.} 

The observed UV luminosity function can be computed from the HMF as
\be
    \Phi_{\rm UV}(L_{\rm obs}) = \int \td L  \td \ln M \frac{\td P(L_{\rm obs}|L)}{\td L_{\rm obs}} \frac{\td P(L|M)}{\td L} \frac{\td n}{\td \ln M}\,  ,
\ee
where $\td P(L|M)/\td L$ denotes the distribution of luminosities emitted by galaxies in halos of mass $M$ and $\td P(L_{\rm obs}|L)/\td L_{\rm obs}$ is the distribution of observed luminosities if the emitted luminosity is $L$. The latter is affected by gravitational lensing magnification~\citep{Takahashi:2011qd} and dust attenuation~\citep{1996ApJ...457..645W}:
\be
    \frac{\td P(L_{\rm obs}|L)}{\td L_{\rm obs}} \approx \int \td \mu \td B \frac{\td P(\mu)}{\td \mu} \frac{\td P(B)}{\td B} \delta\!\left(L_{\rm obs} - \frac{\mu}{B} L\right) \,,
\ee
where $\td P(B)/\td B$ denotes the distribution of the dust attenuation $B$ and $\td P(\mu)/\td \mu$ the distribution of the lensing magnification $\mu$.

\subsection{Lensing magnification}

The first correction to the observed luminosity that we consider is gravitational lensing. This correction is relevant for the bright end of the UV luminosity function, which receives contributions from magnified distant galaxies lensed by intervening structures. As shown by~\citet{Takahashi:2011qd}, the magnification can be approximated by\footnote{This approximation underestimates the high-$\mu$ tail of the distribution. However, this tail is damped by the finite sizes of the lenses.}
\be \label{eq:mu}
    \mu \approx (1-\kappa)^{-2} \,, 
\ee
where $\kappa$ denotes the convergence given by the Laplacian of the lens potential. For multiple lenses, the convergence is obtained by summing the contributions from the different lenses $j$ and subtracting the empty beam convergence $\kappa_{\rm E}$:
\be \label{eq:kappa}
    \kappa = \kappa_{\rm E} + \sum_j \kappa^{(1)}(\vect{\theta}_j) \,, \quad \kappa^{(1)}(\vect{\theta}_j) = \frac{8\pi D_{j} D_{j,s}}{D_s} \Sigma_j(r_j) \,,
\ee
where $\Sigma_j(r_j)$ denotes the projected surface mass density of the $j$th lens, $D_{j}$ the angular diameter distance of the lens, $D_s$ the angular diameter distance of the source and $D_{j,s}$ the angular diameter distance between the lens and the source. The convergence of the $j$th lens depends on the mass of the lens $M_j$, the lens redshift $z_j$ and the distance of the lens from the line of sight to the source $r_j$, $\vect{\theta}_j = \{M_j, z_j, r_j\}$. As mean convergence is $\langle\kappa\rangle = 0$, the empty beam convergence can be written as $\kappa_{\rm E} = -\langle\sum_j \kappa^{(1)}(\vect{\theta}_j) \rangle$. 

\begin{figure}
    \centering
    \includegraphics[width=0.9\columnwidth]{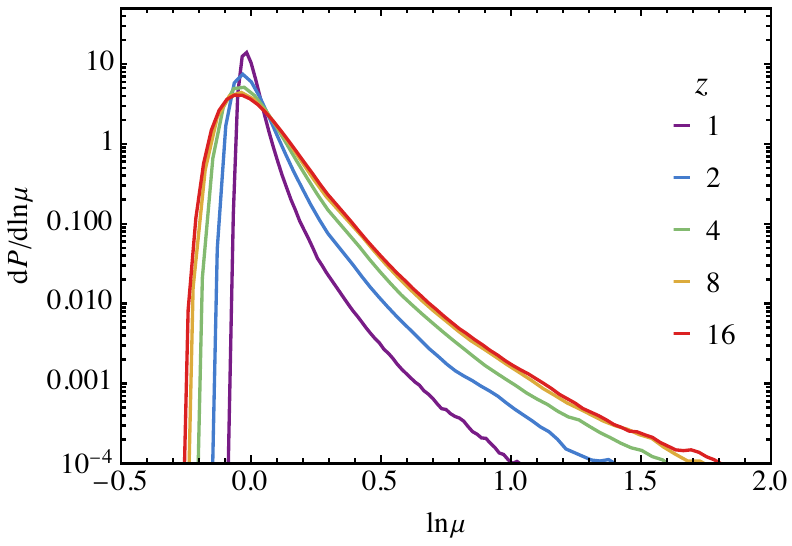}
    \caption{The distributions of lensing magnifications at different source redshifts.}
    \label{fig:lensing}
\end{figure}

We compute the distribution of $\mu$ by generating numerically the distribution of $\kappa$, converting it to $\td P(\mu)/\td \mu$ using Eq.~\eqref{eq:mu} and applying a $1/\mu$ factor to transform the image-plane distribution into the source-plane distribution. We model the lenses by NFW halos whose spatial distribution follows a uniform distribution and use the halo mass-concentration/redshift relation derived by~\citet{Ludlow:2016ifl}. We account for the finite source sizes by averaging the surface density over a sphere of radius $R_s$ that we obtain by projecting the source size to the lens plane. We fix the source size to $10\,{\rm kpc}$ (see in~\citet{Takahashi:2011qd,Ferrami:2022xfy} how the result is affected by the source size). The distribution of lenses is given by
\be
    \frac{\td N(\vect{\theta})}{\td \ln M} 
    = \int_0^{z_s} \td z_l \int_0^{r_{\rm max}} \td r \, \frac{2\pi (1+z_l)^2 r}{H(z_l)} \frac{\td n(z_l)}{\td \ln M} \,.
\ee
To limit the number of lenses $N$, we impose an upper limit on the distance $r$ of the halo from the line of sight, $r < r_{\rm max}(M,z_l)$, so that $\kappa^{(1)} > \kappa_{\rm thr}$. We have checked that our choice of $\kappa_{\rm thr}$ is sufficiently small that our results are insensitive to it. The distribution of contributions $\kappa^{(1)}$ is given by
\bea
    &\frac{\td P^{(1)}(\kappa)}{\td \kappa} = \frac1N \int \td N(\vect{\theta}) \,\delta(\kappa - \kappa^{(1)}(\vect{\theta})) \\
    &\,\,= \frac1N \int \td z_l \td r \left| \frac{\td \kappa^{(1)}}{\td \ln M} \right|^{-1} \frac{\td N(\vect{\theta})}{\td \ln M \td z_l \td r}\bigg|_{M: \,\kappa^{(1)} = \kappa} \,. 
\eea
We compute the distribution of the convergence by picking random numbers from $\td P^{(1)}(\kappa)/\td \kappa$ to generate multiple realizations of $\sum_j \kappa^{(1)}(\vect{\theta}_j)$ and shifting the resulting distribution by $\kappa_{\rm E}$ so that its mean is zero.\footnote{A similar approach of finding the distribution of a sum of random variables was introduced by~\citet{Ellis:2023dgf} for the computation of the gravitational wave background from a population of supermassive black hole binaries.}

\begin{figure}
    \centering
    \includegraphics[width=0.9\columnwidth]{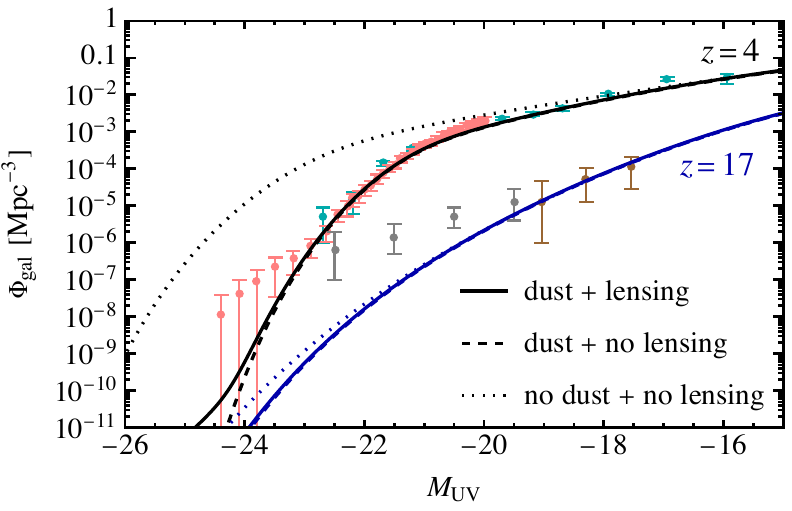}
    \caption{The UV luminosity function at $z=4$ (black) and at $z=17$ (blue) with and without corrections due to lensing magnification and dust attenuation. The cyan and pink points show, respectively, the HST $z=4$ data from~\citet{2021AJ....162...47B} and \citet{2022ApJS..259...20H}, and the brown and gray points the JWST $z=17$ data from~\citet{Perez-Gonzalez:2025bqr} and \citet{Castellano:2025vkt}.}
    \label{fig:lensinganddust}
\end{figure}

The resulting distributions of lensing magnifications at different source redshifts are shown in Fig.~\ref{fig:lensing}. The distributions have a characteristic long tail towards high magnifications. This amplifies the high-luminosity (low magnitude) tail of the UV luminosity functions, as seen in Fig.~\ref{fig:lensinganddust} by comparing the dashed and solid curves. We note that the lensing amplification is dominated by relatively heavy structures, $M\gtrsim 10^{10}\,\Msun$, and is not significantly affected by the changes in the small-scale structures that we consider in this work. 

\subsection{Dust attenuation}

Following the literature (see, e.g.,~\citet{Trenti:2014hka,2020MNRAS.492.5167V}), we neglect the variance in the dust attenuation $B$ and approximate $\td P(B)/\td B \approx \delta(B - \bar B)$. This gives
\be
    \Phi_{\rm UV}(L_{\rm obs}) 
    = \int \td \mu\, \td \ln M \frac{\bar{B}}{\mu} \frac{\td P(\mu)}{\td \mu} \frac{\td P(L|M)}{\td L} \frac{\td n}{\td \ln M} \,,
\ee
where $L = \bar{B}L_{\rm obs}/\mu$.

The UV luminosity functions are commonly presented in terms of the absolute magnitude instead of the luminosity. The associated absolute magnitude $M_{\rm UV}$ is defined by~\citep{Oke:1983nt}
\be
    \log_{10}\left[\frac{L_{\rm UV}}{{\rm erg\,s^{-1}}}\right] = 0.4 \left(51.63-M_{\rm UV}\right)\, ,
\ee
where the normalisation is chosen to match the luminosity of the star Vega. Dust attenuation shifts the absolute magnitude by $A_{\rm UV} = 2.5 \log_{10} \bar{B}$, which we estimate as
\be \label{eq:AUV}
    A_{\rm UV} = \frac1s \ln\left[e^{s(C_0 + 0.2\ln 10 (C_1 \sigma_\beta)^2 + C_1 \bar\beta_{\rm UV})} + 1 \right]
\ee
with $s=3$, $C_0 = 4.4$, $C_1 = 2.0$, $\sigma_\beta = 0.34$ and 
\be
    \bar\beta_{\rm UV} = -\exp\!\left[\frac{0.17(19.5+M_{\rm UV})}{1.54+0.075 z}\right] (1.54+0.075 z) \,.
\ee
The latter is obtained by fitting the results shown in Table 3 of~\citet{Bouwens:2013hxa}, assuming that $\bar\beta_{\rm UV}$ scales exponentially with $M_{\rm UV}$. Notice that, in Eq.~\eqref{eq:AUV}, instead a piecewise function, we use a softplus function with sharpness parameter $s=3$ chosen so that the resulting dust attenuation roughly matches that obtained by numerical sampling shown in Fig.~1 of~\citet{2020MNRAS.492.5167V}. 

In terms of absolute magnitude, the UV luminosity function is given by
\bea
    &\Phi_{\rm UV}(M_{\rm obs}) = \Phi_{\rm UV}(L_{\rm obs}) \frac{\td L_{\rm obs}}{\td M_{\rm obs}} \\
    &\quad = \int \td \mu \frac{\td P(\mu)}{\td \mu} \int \td \ln M \frac{\td n}{\td \ln M} \frac{\td P(M_{\rm UV}|M)}{\td M_{\rm UV}} \,,
\eea
where $M_{\rm UV} = M_{\rm obs} - A_{\rm UV} + 1.086 \ln \mu$. Comparing the short- and long-dashed curves in Fig.~\ref{fig:lensinganddust}, we see that dust attenuation suppresses strongly the high-luminosity tail of the luminosity function at low redshifts but at very high redshifts its effect is almost negligible. However, we note that the dust attenuation is calibrated to observations at $z \le 8$~\citep{Bouwens:2013hxa} and we have simply extrapolated these results to $z\gg 8$.

\subsection{Star formation rate}

\begin{figure}
    \centering
    \includegraphics[width=\columnwidth]{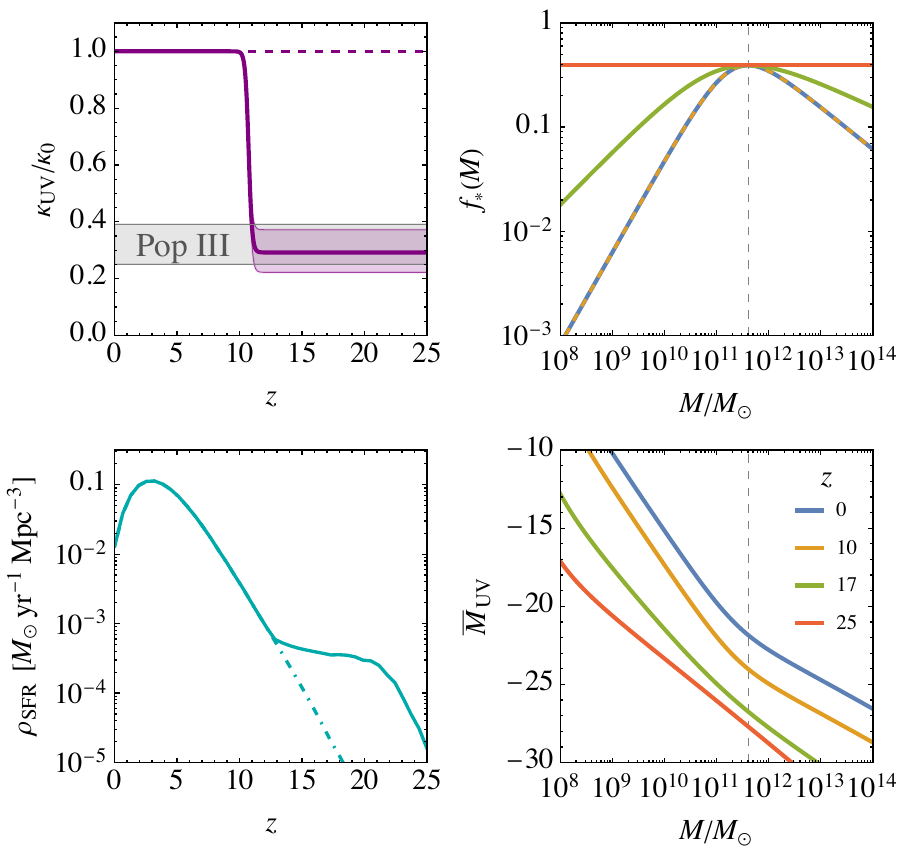}
    \caption{\emph{Upper left:} The thick purple curve shows the best-fit UV conversion factor~\eqref{eq:kappaUV} as a function of redshift and the purple band reflects the 68\% range in $f_{\kappa}$. \emph{Upper right:} The best-fit SFR~\eqref{eq:fstar} as a function of the halo mass at different redshifts. \emph{Lower left:} The SFR density with (solid) and without (dot-dashed) the suppression~\eqref{eq:feedback} of the feedback effects. \emph{Lower right:} The best-fit mean UV magnitude as a function of the halo mass at different redshifts.}
    \label{fig:kappa}
\end{figure}

We assume that the emitted UV magnitudes $M_{\rm UV}$ in a given halo of mass $M$ follow a Gaussian distribution:
\be \label{eq:PMUV}
    \frac{\td P(M_{\rm UV}|M)}{\td M_{\rm UV}} = \frac{1}{\sqrt{2\pi} \sigma_{\rm UV}} \exp\left[ -\frac{(M_{\rm UV} - \bar{M}_{\rm UV})^2}{2\sigma_{\rm UV}^2} \right] \,.
\ee
The scatter in the $M_{\rm UV}-M$ relation, $\sigma_{\rm UV}$, is taken as a free parameter. The mean emitted UV luminosity, $\bar{M}_{\rm UV}$, is directly proportional to the SFR:
\be
    \bar{L}_{\rm UV} = \frac{\dot{M}_*}{\kappa_{\rm UV}} \,.
\ee
As mentioned in the Introduction, we assume that the SFR is proportional to the halo growth rate~\citep{2013ApJ...774...28B}, $\dot M_* = f_B f_*(M) \,\dot M$, where $f_B = \Omega_B/\Omega_M \approx 0.16$. We parametrise the proportionality coefficient $f_*(M)$, which models the feedback effects from  SNe and AGNs, as a broken power-law around $M = M_c$ with an exponential suppression in halos with $M < M_t < M_c$:
\be \label{eq:fstar}
    f_*(M) = \epsilon \,\frac{\alpha+\beta}{\beta(M/M_c)^{-\alpha}+\alpha(M/M_c)^{\beta}} \,e^{-M_t/M} \,,
\ee
where $\alpha,\, \beta,\, \epsilon > 0$. {A similar parametrization is used, e.g., in the \texttt{GALLUMI} code~\citep{Sabti:2021xvh}. The difference is that, in our parametrisation, the maximum of $f_*(M)$ is at $M = M_c$ for $M_t \ll M_c$.} We find that $z$-independent values of $\alpha$, $\beta$, $\epsilon$ and $\kappa_{\rm UV}$ give a good fit to the observations up to $z\approx 10$, but that the data prefer higher luminosities at $z\gtrsim 10$. 

To accommodate this enhancement in the luminosities of the early stars, we introduce a parametrisation in which the conversion factor $\kappa_{\rm UV}$ changes around $z=z_\kappa$: 
\be \label{eq:kappaUV}
    \kappa_{\rm UV} = \kappa_0 \left[\frac{1 + f_\kappa}{2} - \frac{1 - f_\kappa}{2} \tanh\left(\frac{z - z_\kappa}{\gamma}\right) \right] \,,
\ee 
where $\kappa_0 = 1.15\times 10^{-22} \Msun \,{\rm s}\, {\rm erg}^{-1} \,{\rm Myr}^{-1}$, $0<f_\kappa\leq 1$ and $\gamma >0$. The conversion factor depends on the initial stellar mass function and the value of $\kappa_0$ corresponds to the Salpeter initial mass function~\citep{Madau:2014bja}. We allow for a wide range of possibilities in the transition to Pop-III stars; $z_{\kappa}$ determines the redshift at which the change of $\kappa_{\rm UV}$ starts, $\gamma$ parametrises how sharp the transition is and $f_\kappa$ the fractional change in $\kappa$, such that at $z \gg z_\kappa$, $\kappa_{\rm UV} = f_\kappa \kappa_0$. We show in the upper left panel of Fig.~\ref{fig:kappa} $\kappa_{\rm UV}$ as a function of $z$ for the best fit values in the CDM model.

In order to fit the highest-$z$ UV data, we find that the feedback effects, which we parametrise with $f_*(M)$, need to be reduced. To allow for such a reduction, we parametrise the powers $X = \alpha,\beta$ so that they can approach zero linearly with $z$ above some redshift $z_{\rm fb}$: 
\be \label{eq:feedback}
    X_{z>z_{\rm fb}} = X \max\left[0, \frac{z_*-z}{z_*-z_{\rm fb}}\right] \,.
\ee
In this parametrisation, $\alpha$ and $\beta$ become zero at $z>z_*$, and $f_*(M)$ is mass independent at $M\gg M_t$, as shown in the upper right panel of Fig.~\ref{fig:kappa} for the best fit values. This implies that in all halos with $M\gg M_t$, the baryons that the halo accretes are converted into stars with efficiency $\epsilon$. 

In total, our parametrization of the SFR, the UV conversion factor and the distribution of the UV magnitudes includes 11 free parameters: $\{M_t,\,M_c,\,\epsilon,\,\alpha,\,\beta,\,\gamma,\,f_\kappa,\,z_\kappa,\,z_{\rm fb},\,z_*,\,\sigma_{\rm UV}\}$.

\section{Results}

We consider UV luminosity function measurements derived from both HST observations~\citep{2021AJ....162...47B, 2022ApJS..259...20H} and JWST observations~\citep{2024MNRAS.533.3222D, Perez-Gonzalez:2025bqr, Castellano:2025vkt}. The latter are based on high-$z$ photometric measurements, most of which are not spectroscopically confirmed. This can lead to errors in the estimation of redshift. Most notably, a photometric measurement of CEERS-93316 indicated that this galaxy was at $z \approx 16$~\citep{2022arXiv220802794N} but a later spectroscopic measurement revealed it to be at $z \approx 5$~\citep{2023Natur.622..707A}. However, the redshifts of some of the high-$z$ galaxies have been confirmed spectroscopically by~\cite{2024MNRAS.533.3222D}. 

We perform a Markov-chain Monte Carlo (MCMC) analysis of the models. For each scan over the model parameters, we generate 8 MCMC chains, each consisting of 12000 samples of which the first 2000 are discarded as burn-in. We use the Metropolis-Hastings sampler with Gaussian proposal distributions whose widths we choose so that the acceptance rate is around $10\%$. The model parameter priors are given in Table~\ref{table:priors}. We assume that the data follow a split normal distribution $\mathcal{N}(\mu,\sigma_1,\sigma_2)$, where $\mu$ denotes the mode and $\sigma_{1,2}$ the left- and right-hand-side standard deviations. The likelihood is given by
\be
    \mathcal{L} = \prod_j \mathcal{N}(\Phi(M_{\rm UV}^{(j)},z^{(j)})-\Phi^{(j)},\sigma_1^{(j)},\sigma_2^{(j)}) \,.
\ee
where $j$ labels the measurements and the function $\Phi(M_{\rm UV},z)$ is the model prediction. The code used in the analysis is available at~\href{https://github.com/vianvask/halos}{GitHub}.

\begin{table}
    \centering
    \caption{Parameters range and best fit values}
    \begin{tabular}{c|c|c}
        \hline
        \hline
        Parameter & Prior & Best fit \\
        \hline
        \hline \\[-9pt]
        $M_t/\Msun$ & LogU$(6,10)$ & $10^{7.9}$ \\
        $M_c/10^{11}\Msun$ & U$(3,5)$ & 4.0 \\
        $\epsilon$ & U$(0.36, 0.42)$ & 0.39 \\
        $\alpha$ & U$(0.6, 1.1)$ & 0.88 \\
        $\beta$ & U$(0.2, 0.6)$ & 0.40 \\
        $\gamma$ & U$(0.05, 0.6)$ & 0.27 \\
        $f_\kappa$ & U$(0.05,  0.7)$ & 0.29 \\
        $z_\kappa$ & U$(9.8, 11.4)$ & 10.7 \\
        $z_{\rm fb}$ & U$(7,25)$ & 12.7 \\
        $z_*$ & U$(20,36)$ & 22.9 \\
        $\sigma_{\rm UV}$ & U$(0.05, 0.2)$ & 0.068 \\
        \hline
        \hline
    \end{tabular}
    \vspace{2mm}
    \tablefoot{Prior ranges and best-fit values of the parameters in the CDM model. The parametrization is given by Eqs.~\eqref{eq:PMUV}, \eqref{eq:fstar}, \eqref{eq:kappaUV} and \eqref{eq:feedback}.}
    \label{table:priors}
\end{table}

\begin{figure*}
    \centering
    \includegraphics[width=\textwidth]{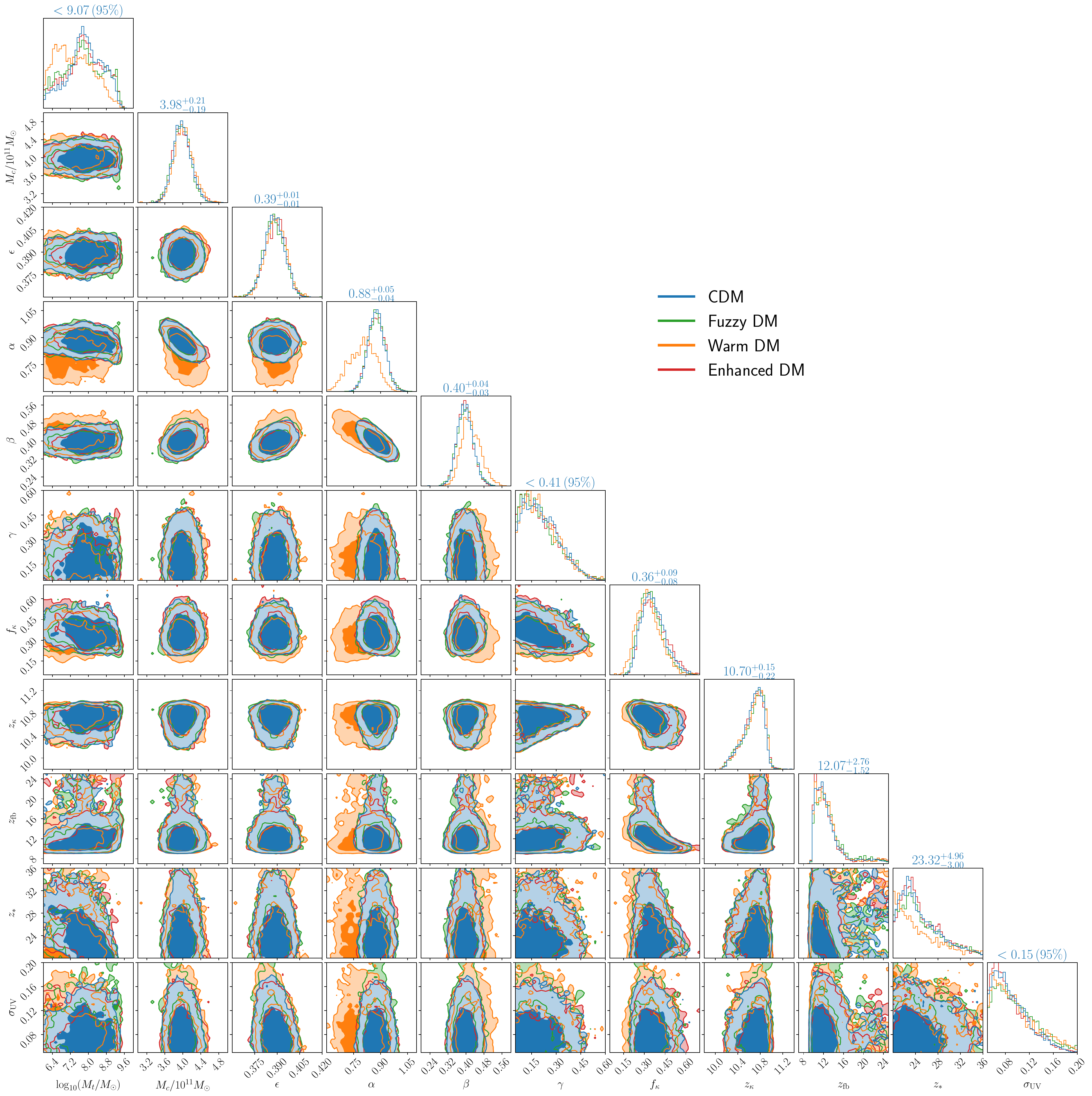}
    \vspace{-2mm}
    \caption{Posteriors of SFR fits to the UV luminosity function observations in different models. The contours indicate the 68\% and 95\% credible regions. The marginalised 68\% credible ranges and the 95\% credible constraints are shown for CDM on top of each column.}
    \label{fig:corner4}
\end{figure*}

\subsection{Cold dark matter} 

The full posteriors for our CDM fit parameters are shown in blue in Fig.~\ref{fig:corner4} and the best-fit values are given in Table~\ref{table:priors}. We have checked that the Gelman-Rubin statistic $R$~\citep{Gelman:1992zz} is very close to $1$ for each of the parameters. This indicates good convergence of the MCMC chains. The highest value, $R\approx 1.09$, was obtained for $\log_{10}(M_t/M_\odot)$.

Most of the parameters, in particular those parametrising the broken power-law shape of $f_*(M)$, are well constrained by the data, and the posteriors show only a mild negative correlation between $\alpha$ and $\beta$ as well as between $\alpha$ and $M_c$. {The fits of $\alpha$, $\beta$ and $M_c$ are similar to those found by~\citet{2022ApJS..259...20H}, but we find about a factor of two higher SFR. Compared to~\citet{Sabti:2021xvh}, we have fixed the cosmological parameters to the Planck CMB values and we are able to constrain $M_c$ and $\beta$ better because the HST data we consider extend to higher luminosities.}

From the posteriors of $\gamma$, $f_\kappa$ and $z_\kappa$, we see a clear ($>95\%$) preference for a relatively rapid change in $\kappa_{\rm UV}$ around $z_\kappa = 10-11$ by factor of $\sim 3$. The posterior of $z_{\rm fb}$ shows that the preference for a high-$z$ change in the powers $\alpha$ and $\beta$ is less significant, $<95\%$. This is expected, since the change is driven by the measurements at $z=17$ and $z=25$ that have relatively large uncertainties and $\Phi = 0$ is within the 95\% range. Moreover, we see a correlation between $f_\kappa$ and $z_{\rm bf}$, indicating that the enhancement of the UV luminosity can be partly reduced by changing the powers $\alpha$ and $\beta$. The data prefer a narrow scatter in the UV magnitude, $\sigma_{\rm UV} < 0.15$ at the $95\%$ CL, and give an upper bound on the scale of the exponential SFR suppression, $M_t < 10^9\,\Msun$ at the $95\%$ CL. This implies that the explanation for the UV excess as a consequence of a large constant scatter in the $(M_{\rm UV}\,, M)$ is disfavoured when compared to the Pop-III star hypothesis.~\footnote{However, a mass-dependent scatter that increases at small halo masses could enhance the UV luminosities at high-$z$~\citep{2024ApJ...975..192G}.} This result should be treated with caution, however, since selection effects, which are not considered in the likelihood analysis, could modify the conclusion. 

The evolution of $\kappa_{\rm UV}$ for the best fit is shown in the upper left panel of Fig.~\ref{fig:kappa}. The range shown in gray corresponds to estimates of the conversion factor for Pop-III stars with top-heavy initial mass functions~\citep{Harikane:2022rqt}. We see that the best fit prefers a sharp transition from Pop-I and Pop-II stars with standard Salpeter profiles up to $z\sim10$ to a top-heavy Pop-III population at $z>10.8$. The upper right panel of Fig.~\ref{fig:kappa} illustrates how the feedback effects are suppressed for the best fit. This strongly enhances the SFR density, 
\be
    \rho_{\rm SFR} = \int \td M \frac{\td n}{\td M} \dot{M}_*(M) \,,
\ee
at high $z$, as shown in the lower left panel of Fig.~\ref{fig:kappa}. The lower right panel of Fig.~\ref{fig:kappa} shows the relation between the mean UV magnitude and the halo mass at different redhifts. 

\begin{figure*}
    \centering
    \includegraphics[width=\textwidth]{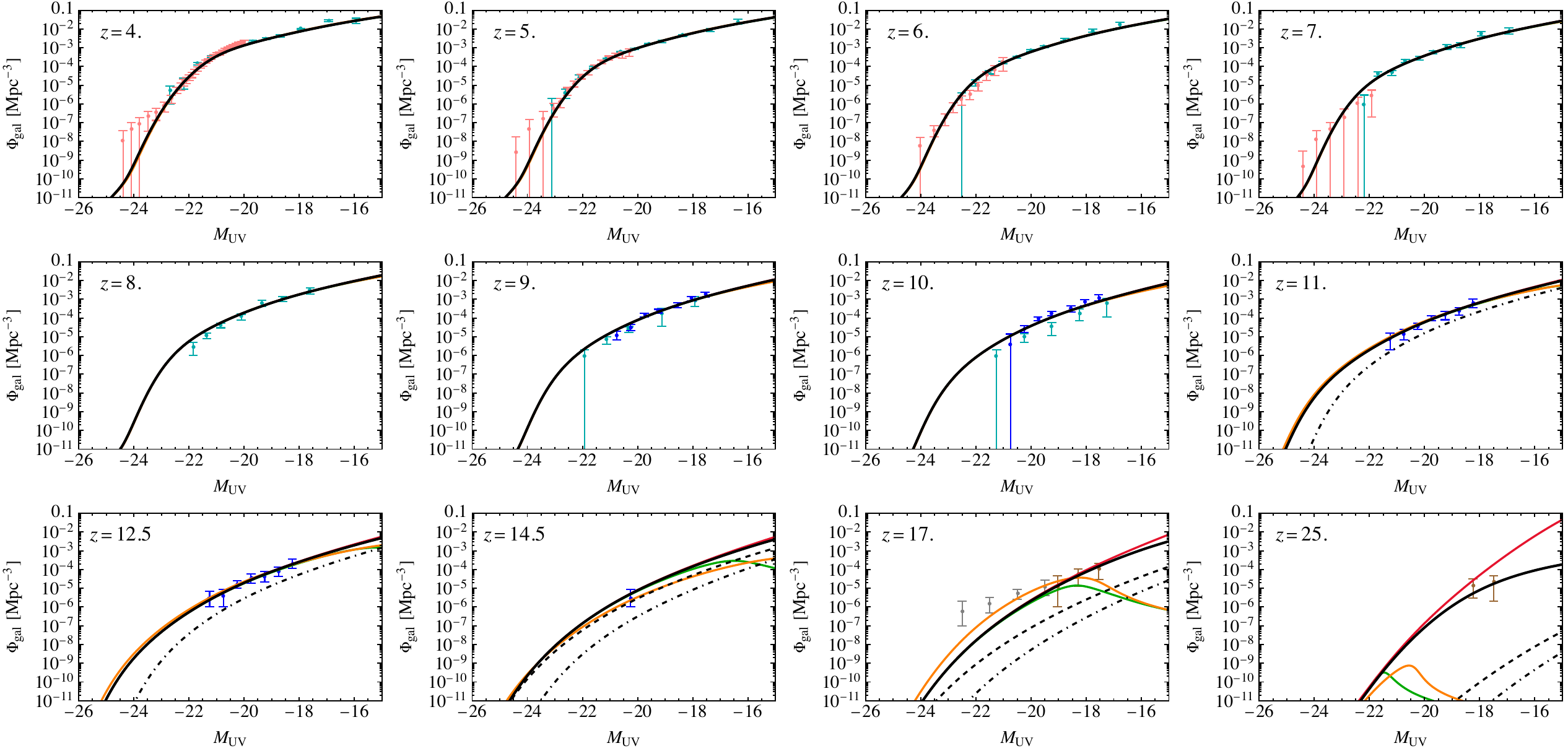}
    \caption{The UV luminosity function at different redshifts in the cold (black), fuzzy (green) and warm (orange) DM models and in the model with enhanced matter power (red). The best fit is shown for CDM and WDM cases, while for FDM the mass corresponds to the 95\% lower bound. The dashed and dot-dashed curves show the CDM UV luminosity function without the suppression of the feedback effects and without the change in the UV conversion factor $\kappa_{\rm UV}$. The points with errorbars show the HST measurements from~\citet{2021AJ....162...47B} (cyan) and \citet{2022ApJS..259...20H} (pink) and the JWST measurements from~\cite{2024MNRAS.533.3222D} (blue), \citet{Perez-Gonzalez:2025bqr} (brown) and \citet{Castellano:2025vkt} (gray).}
    \label{fig:UVLF}
\end{figure*}

In Fig.~\ref{fig:UVLF} we show the best-fit UV luminosity function in the CDM model (in black) compared with the HST and JWST observations at different redshifts. The change in the tail shape at $M_{\rm UV}<-22$ is primarily driven by the lensing magnification. The slight turn seen in the last panel at $M_{\rm UV} > -18$ is due to the exponential term in~\eqref{eq:fstar} that becomes relevant in the shown $M_{\rm UV}$ range only at the highest $z$, as seen from the lower right panel of Fig.~\ref{fig:kappa}. The fit without the suppression of the feedback effects is shown by the dashed lines, and undershoots the measurements at $z=17$ and $25$. The dot-dashed curve shows the fit when also the high-$z$ enhancement in the luminosities is also removed. We see that this case undershoots the observed luminosity function at redshifts $z>10$. Moreover, we see that our model does not provide a good fit of the brightest objects at $z=17$. This may reflect a limitation of our parametrisation, but it is also possible that the luminosities of some of these objects are contaminated by AGNs~\citep{Castellano:2025vkt}.

\begin{figure*}
    \centering
    \includegraphics[width=\textwidth]{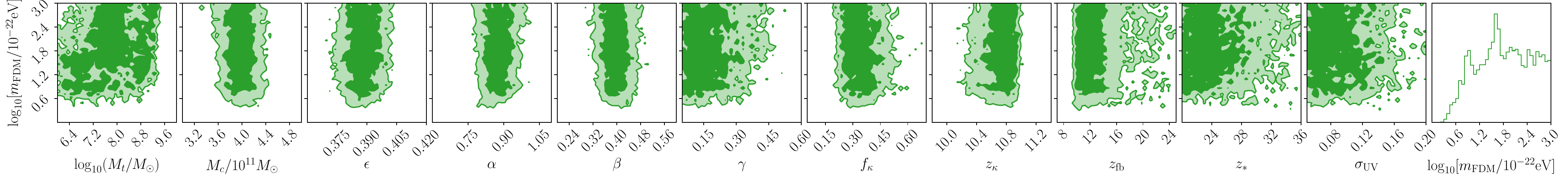}
    \includegraphics[width=\textwidth]{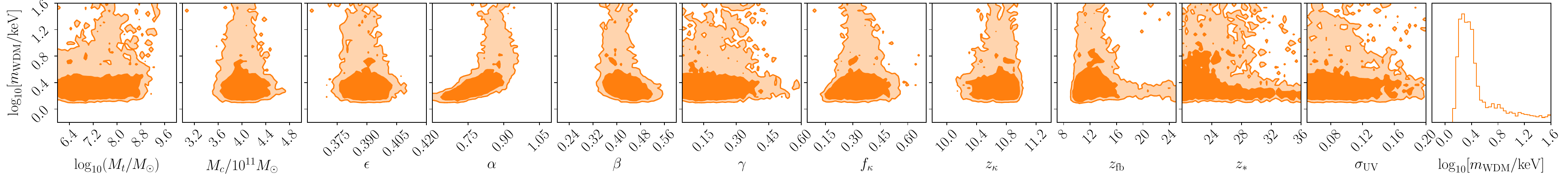}
    \includegraphics[width=\textwidth]{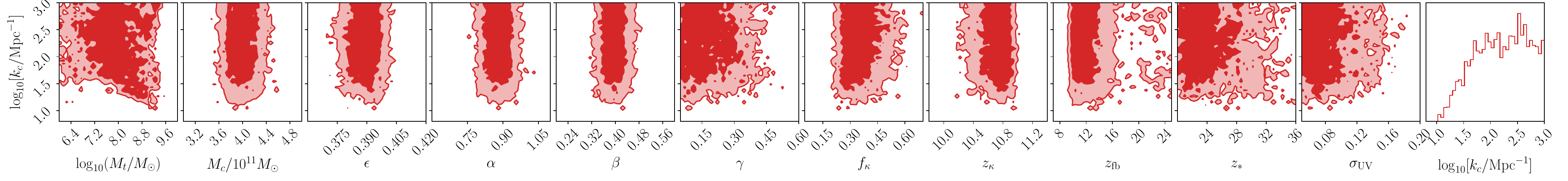}
    \vspace{-3mm}
    \caption{Posteriors of SFR fits to the UV luminosity function observations as a function of FDM mass (top panels), WDM mass (middle panels) and the enhancement scale (bottom panels). The contours indicate the 68\% and 95\% credible regions.}
    \label{fig:marg}
\end{figure*}

\subsection{Suppressed small scale perturbations} 

We also see in Fig.~\ref{fig:UVLF} the comparison between the UV data and the predictions with suppressed small-scale structures in the FDM (green) and WDM (orange) models. For FDM we show the case where the model parameters correspond to the CDM best fit and the FDM mass is at the 95\% CL lower limit, while for WDM we show the best fit case, which is slightly better than in the CDM model. We see that the predictions are indistinguishable for $z \lesssim 10$, and up to $z\sim15$ the suppression is at lighter halos (bigger $M_{\rm UV}$) than what the observations can probe. At the highest redshifts, $z>15$, we see that the model predictions undershoot the observations. However, since the uncertainties in the data are large, the observations at high-$z$ are compatible with a strongly suppressed UV luminosity function.

For the FDM model, the parameters for the star formation rate, the enhancement of the UV luminosity and the suppression of the feedback effects remain the same as in the CDM model, as seen in the posterior plots in Fig.~\ref{fig:corner4}. This reflects the fact that the FDM prediction deviates sharply from the CDM prediction at the suppression scale. The main difference between FDM and CDM models is seen in $z_*$, which determines when the feedback effects are fully removed. For small FDM masses the feedback effects can be fully removed at lower $z$ because the luminosity function is suppressed by the FDM quantum pressure, which causes the peak in the posterior of the FDM mass shown in the top panels of Fig.~\ref{fig:marg}. The posterior also shows a lower bound on the FDM mass, $m_{\rm FDM}> 5.6\times10^{-22}\,{\rm eV}$ at 95\% CL. The UV luminosity function has been previously studied in the FDM model in~\citep{Bozek:2014uqa,Schive:2015kza,Corasaniti:2016epp,Winch:2024mrt,Lazare:2024uvj}. Our constraint on the FDM mass is somewhat stronger than that obtained by~\citet{Winch:2024mrt} using the HST data and JWST data from~\cite{2024ApJ...960...56H}.

The deviation from the CDM prediction is smoother in the WDM case, and starts already much above the half-mode mass. Consequently, the WDM fit at small $m_{\rm WDM}$ moves to smaller values of $\alpha$ and larger values of $\beta$ than in the CDM model, as seen in the posterior plots in Figs.~\ref{fig:corner4} and~\ref{fig:marg}. Due to the degeneracy between $\alpha$ and $m_{\rm WDM}$, in particular, the WDM model favours a narrow mass range around 2\,keV. The maximal likelihood in the WDM model is slightly higher than in CDM, $\Delta \ln \mathcal{L} \approx 2.7$, and the Bayesian Information Criterion (BIC), which accounts for the increase in the number of parameters, indicates a mild (insignificant) preference for WDM over CDM with $\Delta {\rm BIC} \approx -0.4$. As seen from Fig.~\ref{fig:UVLF}, the fit improves in particular at $z = 17$.

The posteriors of the WDM mass shown in the middle panels of Fig.~\ref{fig:marg} show a lower bound $m_{\rm WDM}> 1.5\,{\rm keV}$ at 95\% CL. Several previous works have derived constraints on WDM from UV luminosity function observations~\citep{Menci:2016eui,Corasaniti:2016epp,Rudakovskyi:2021jyf,Hibbard:2022sng,Maio:2022lzg,Liu:2024edl}. Although we use a larger set of JWST data, the bound on the WDM mass is weaker than what was found by~\citet{Liu:2024edl}. This difference arises because~\citet{Liu:2024edl} uses a low-$z$ fit of the HMF to numerical simulations derived in~\citet{Stucker:2021vyx}, while we estimate the WDM HMF using the transfer function and the smooth-$k$ window function. 

\subsection{Enhanced small-scale perturbations} 

\begin{figure}
    \centering
    \includegraphics[width=0.98\columnwidth]{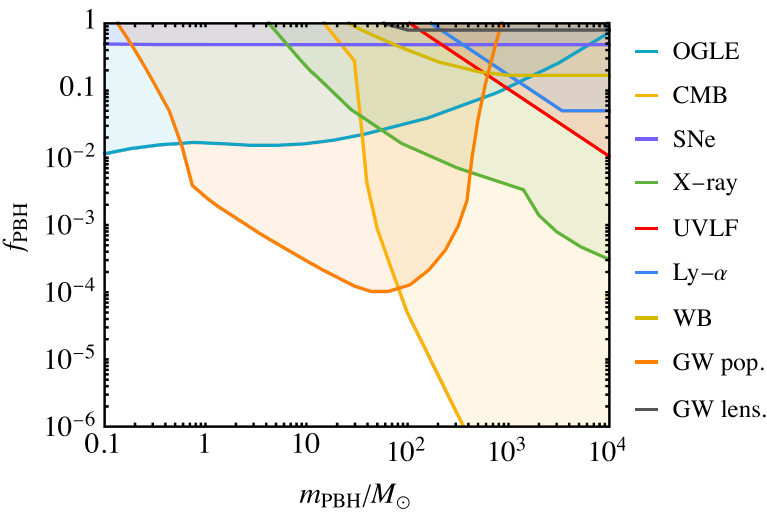}
    \caption{Constraints on the fraction of the DM density provided by PBHs from the OGLE microlensing survey~\citep{Mroz:2024mse}, accretion limits from CMB observations~\citep{Serpico:2020ehh}, SNe lensing~\citep{Zumalacarregui:2017qqd}, the population of X-ray sources~\citep{Inoue:2017csr}, Lyman-$\alpha$ data~\citep{Murgia:2019duy}, survival of wide binaries~\citep{Monroy-Rodriguez:2014ula}, the population of gravitational wave (GW) sources~\citep{Andres-Carcasona:2024wqk} and GW lensing~\citep{Urrutia:2023mtk}. The red line shows the upper limit derived in this work from measurements of the UV luminosity function.}
    \label{fig:PBHs}
\end{figure}

Enhancing the small-scale structure has the opposite effect as WDM or FDM, but the conclusions are very similar. We see in Fig.~\ref{fig:UVLF} that for $z\lesssim10$ the predictions resemble those of CDM, and only at the highest redshifts do we see a small enhancement of the UV luminosity function that grows at lower scales, $M_{\rm UV}\lesssim-20$. This reflects the fact that the higher the redshift, the more deviations from CDM one can expect to see, especially for low-mass halos. When fitting the star formation parameters, the results are independent of the enhancement of the scales and very similar to CDM, as seen in Fig.~\ref{fig:corner4}. We conclude that enhancing the matter power spectrum does not help to explain the excess in the UV radiation at high-$z$ when other possibilities like Pop-III stars are taken into account. The fit favours a sharp transition to Pop-III stars and such a feature cannot be reproduced by enhancing the matter power spectrum. The fit gives a lower bound $k_c>25\,{\rm Mpc}^{-1}$ at 95\% CL, as seen in the bottom panels of Fig.~\ref{fig:marg}. The excluded region corresponds to axion miniclusters produced with $m_a < 6.6\times10^{-17}\,{\rm eV}$ or PBHs that obey $f_{\rm PBH} > \max[105 \Msun/m_{\rm PBH}, 10^{-4} (m_{\rm PBH}/10^4 \Msun)^{-0.09}]$ where the second bound comes from the requirement that $k_c < k_{\rm cut}$. Our constraint on $k_c$ is significantly stronger than that implied by the measurements of the matter power spectrum at scales $0.5 < k/{\rm Mpc}^{-1} < 10$ derived by~\citet{Sabti:2021unj} and that implied by the analysis of~\citet{Irsic:2019iff}, both using the HST data alone. As for PBHs, our constraint is stronger than that derived by~\citet{Murgia:2019duy}, which considered the limit on an enhanced small-scale power spectrum imposed by the Lyman-$\alpha$ forest observations, and that derived by~\citet{Gouttenoire:2023nzr} using the HST UV luminosity function data. In Fig.~\ref{fig:PBHs}, we show a compilation of previous PBH constraints together with our new constraint in red.

\section{Conclusions}

We have investigated in this work the implications of the high-$z$ UV luminosity function observations with HST and JWST for star formation and different DM models. We have revisited the excursion-set formalism to compute the unconditional and conditional halo mass functions and derived a new estimate of the halo growth rate. We have also proposed a new window function, inspired by the smooth-$k$~\cite{Leo:2018odn} window function, that reproduces well the numerical results and matches better the CDM predictions at high masses. 

We have found that, assuming that the SFR is proportional to the halo growth rate, the observations of the UV luminosity function indicate two interesting effects that change the luminosity function at high $z$. First, we find evidence at more than $95\%$ CL for a sharp enhancement in the luminosity function around $z \simeq 10$. This can be explained by a transition from a Salpeter mass function to a top-heavy Pop-III mass function. Beyond $z=10-15$ the luminosity function needs to be further enhanced, in particular to match the JWST observations at $z=17$ and 25. This enhancement can be achieved by reducing the feedback from SNe and AGNs. The evidence for this second effect is, however, less than $95\%$ CL. We have also considered the effect of dust attenuation and lensing, and shown that they are relevant for the bright end of the luminosity function at $z<10$ but do not play a significant role in fitting the higher redshift data. 

We have considered departures from CDM in two directions, suppressing the small-scale structures via either warm or fuzzy DM and enhancing the matter power spectrum with white noise, which could happen in scenarios including axion miniclusters or PBHs. When marginalising over the SFR parametrisation, we find a lower bound of $5.6 \times 10^{-22}{\rm eV}$ for the FDM mass and $1.5\, {\rm keV}$ for the WDM mass, at $95\%$ CL. Because the new JWST data at high-$z$ still have significant uncertainties, we find that the DM bounds are not much stronger than those obtained using the HST data alone. Another important conclusion is that enhanced matter perturbations are not preferred over changes in the star formation rate due to the presence of Pop-III stars and changes in the feedback in the pre-reionisation era. We have found that the data constrain the scale of the enhancement to $k_c > 25\,{\rm Mpc}^{-1}$ at $95\%$ CL, excluding, for example, the possibility of axion miniclusters produced for $m_a < 6.6 \times 10^{-17}\,{\rm eV}$ or heavy PBHs that constitute a fraction $f_{\rm PBH} > \max[105 \Msun/m_{\rm PBH}, 10^{-4} (m_{\rm PBH}/10^4 \Msun)^{-0.09}]$ of DM. 

As JWST gathers more data and the uncertainties are reduced, the bounds on DM will get significantly stronger and our knowledge of star formation at high-$z$ will improve. 

In a forthcoming paper we will leverage our findings to compute the growth of SMBHs in non-CDM scenarios and obtain more competitive bounds on deviations from CDM using the measurements of high-$z$ SMBHs by JWST. 

\begin{acknowledgements}
This work was supported by the Estonian Research Council grants PRG803, PSG869, RVTT3 and RVTT7 and the Center of Excellence program TK202. The work of J.E. and M.F. was supported by the United Kingdom STFC Grants ST/T000759/1 and ST/X000753/1.  The work of V.V. was partially supported by the European Union's Horizon Europe research and innovation program under the Marie Sk\l{}odowska-Curie grant agreement No. 101065736.
\end{acknowledgements}

\bibliographystyle{aa}
\bibliography{refs}

\end{document}